\documentclass[12pt]{article}
\usepackage{graphicx}

\begin{document}

\def\kk{{\bf k_1}}
\def\kg{{\bf k_2}}
\def\qq{{\bf q}}
\def\pom{{I\!\!P}}

\centerline{\large \bf Describing $F_{2}$ through a finite sum of
gluon ladders}

\vskip 0.7cm

\centerline{M.B. Gay Ducati $^{(a)}$, K. Kontros $^{(b)}$, A. Lengyel $^{(b)}$,
M.V.T. Machado $^{(a)}$}

\vskip 0.3cm

\centerline{\sl $^{(a)}$ Instituto de F\'{\i}sica, Universidade Federal do
Rio Grande do Sul} \centerline{\sl Caixa Postal 15051, CEP
91501-970, Porto Alegre, RS, Brazil}

\centerline{\sl $^{(b)}$ Institute of Electron Physics, National
Academy of Sciences of Ukraine}\centerline{\sl Universitetska 21,
UA-88016 Uzhgorod, Ukraine}

\abstract{Electroproduction in deep inelastic scattering at HERA
is studied in a model considering a finite sum of gluon ladders,
associated with a truncation of the BFKL series. The approach
contains the bare two gluon exchange and both one and two rungs
contributions.  The model is
fitted to the data on the inclusive structure function
$F_2(x,Q^2)$ in the region $x<0.025$ and $0.045<Q^2<1500$
GeV$^2$, with a good agreement. Such a description for a large
span in $Q^2$ is obtained through a suitable modeling of the
remaining non-perturbative background.}


\bigskip

\section{Introduction}

The high energy limit of the photon-proton scattering has been one of the
main open questions concerning perturbative QCD and there is a great
theoretical challenge in  describing such 
process. The successful renormalization group
 approach summing the
contributions of order $(\alpha_s\ln Q^2))^n$, namely the DGLAP evolution
equations \cite{DGLAP}, which has
 described systematically the deep inelastic
data (gluon driven) starts to present slight
 deviations as the energies
reached in the current experiments have increased \cite{THERA}. Although its 
limitation has been theoretically
  determined \cite{OPEbreak}, the DGLAP
approach has enough flexibility  to describe virtual photon initiated
reactions on both low and high
 virtualities at small $x$ \cite{MKlein}.
Despite this fact, non-linear
 effects to the standard DGLAP formalism,
associated with parton (mainly gluons) saturation
 and unitarity corrections
\cite{saturation}, are already known and their
 importance to describe the
relevant observables at small $x$ and estimate
 the future measurements is not
negligible \cite{sat2}.

On the other hand, the high energy QCD calculation encoded in the leading
logarithmic approximation (LLA) BFKL formalism \cite{LOBFKL} summing the
contributions $(\alpha_s \ln (1/x))^n$ is a powerful
 technique to perform
predictions to the physical processes where a clear
 hard scale takes place,
in which the approximation of fixed strong coupling (instead of running) in
some external scale of the process is considered. This evolution equation
provides the $x$ evolution of the gluon density at small $x$. The main
difficulty regarding this approach is the
 resulting total cross section for
the BFKL Pomeron exchange  that violates the
 unitarity limit, stated by the
Froissart limit $\sigma_{tot} \leq \ln^2
 (s)$ \cite{Froissart} ($\sqrt{s}$ is
the center of mass energy), in hadronic collisions. It is a general believe
that this bound is also required in $\gamma^* p$ reactions. Recently, the
next-to-leading (NLLA) BFKL calculation program has been accomplished showing
that the  convergence of the series and its stability is far from clear at the
moment \cite{Salam}. Nevertheless, the
 unitarity problem has been  addressed
in Ref. \cite{Kovchegov}, where the
 multiple LO BFKL Pomeron exchanges
dominate over the next order
 corrections and then the final scattering
amplitude is unitarized. Such an
 approach gives rise to an evolution equation
for the parton densities matching several statements of the saturation
approaches mentioned above.

Recently, we have
proposed a  different phenomenological way to
 calculate the total cross
sections in the hadronic sector
 \cite{Trunkpp}, i.e. proton-(anti)proton
collisions, and in deep
 inelastic scattering \cite{Trunkep}. As a starting
point, in Ref.
 \cite{Fiore} the reliability to describe the $pp(p\bar{p})$
total cross sections through a QCD inspired calculation was shown
 to be
successful. The main point is: at finite total energies, the LLA and NLLA
summation implies that the amplitude is represented by a finite sum on terms,
where the number of terms increases like $\ln s$, rather than by the solution
of the BFKL integral equation. So one perform a truncation in the BFKL
series and use it for a phenomenological description of the non-asymptotic
energy data. The interest in to take the  firts terms in the completete series
in the  truncation is related to the fact that the energies reached by the
present  accelators are not high enough to accommodate a big number of gluons
in the ladder rungs that eventually hadronize. Corrobating this hypothesis,
for example the coefficient weighting the term $\sim \ln^3 s$ turn out to be
compatible with zero considering even the Tevatron data \cite{Fiore}, in
contrast with the expected from the complete resummation. This finding suggest
that energies reached by the present accelerators are not yet asymptotic.

Moreover, in the present theoretical
scenario the NLO corrections to the LO BFKL approach were accomplished, which
lead to a decreasing of the hard Pomeron intercept, issues about the running
coupling constant and modifications in the correspondent anomalous dimension
(for a short review, see \cite{Salam}). Definitive answers concerning the NLO
solution  are now far from clear, imposing severe limits on the possible
applicability of those results to the experimental situation (see, for example
\cite{ABB}).  Some remaining  pathologies has been cured in the 
generalization of the high energy factorization formula in next-to-leading
approximation by Ciafaloni and collaborators \cite{Ciafaloni}. The improving
procedure is the identification of the collinearly-enhanced physical
contributions as the most important agent of the instability of the BFKL
hierarchy.   Although of this reasonably  determined theoretical status, a
comparison of the underlying expressions with the deep inelastic data like in
the case of LO BFKL was not performed yet. Indeed, currently the main focus is
for processes with a predominantly hard scale, for instance $\gamma^*
\gamma^*$ or forward-jets, instead a two scale process like DIS. More
specificallly, in the full resummation of the BFKL series even at LLA there is
not enough flexibility to perform modifications to consider the complete 
subleading contributions. A partial calculation of the  subleading correction
at the LO BFKL has been performed \cite{kw}, taking into account dominant
non-leading effects which come from the requirement that the virtuality of the
exchanged gluons along the gluon ladder is controlled by their transverse
momentum squared, restricting the available phase-space of the emitted gluons.
Such a procedure can be used to perform predictions to observables to be
measured.

Regarding the important question of take into account subleading contributions
in a fixed order expansion of the BFKL approach, this issue was firstly
addressed by Fiore et al. in the hadronic sector \cite{Fiore}. The coeficients
of different powers of $\ln(s)$ in the series refer to the dominant
contribution, at asymptotic energies, for each perturbative order. However,
when we perform a phenomenological description of data the subleading
contributions are absorbed in the normalization constants for each ladder in
the expansion or in the non-perturbative parameters of the model. This is the
main factor of reliability lying in the data description of Refs.
\cite{hepsasha,Fiore} and in our more detailed study for $pp (p\bar{p})$
reactions in Ref. \cite{Trunkpp}.

Motivated by the good result of the truncation in the hadronic case, we
performed a more
 detailed calculation for those observables and extended the
analysis to the non-forward region. Concerning high energy
 limit, the BFKL
approach was a natural choice, since it  takes into
 account a n-rung ladder
contribution. To address a finite phase
 space for the gluon emission,
disregarding the resummation and
 NLO effects intrinsic to the calculations,
we consider that the
 possible exchanged ladders are builded by a little
number of gluon rungs. A
 non-asymptotic  cross section is obtained
considering only the
 one rung contribution ($\sigma_{tot} \sim \ln\, s$),
which is
 enough to describe the available data \cite{Trunkpp}. The non
 zero
momentum transfer calculation was performed and the
 role played by a suitable
choice of the proton impact factor was
 determined. Moreover, an useful
parametrization for the elastic
 slope $B_{el}(s)$ \cite{Trunkpp}, consistent
with the Regge phenomenology, was
 introduced allowing to describe with good
agreement the
 differential cross section. The main features coming from that
calculation are also corroborated by the phenomenological studies
 in Ref.
\cite{hepsasha}.
 
 Focusing on the deep inelastic scattering, the similar
motivation for the truncation of the perturbative series is the
 small phase
space to allow an infinite gluon cascade in the final
 state. Some authors
even advocate  that the  resummation
 technique in deep inelastic is not
completely correct
 \cite{Ryskin}. Indeed, in the available energies there is
no room
 on pseudorapidity to enable a completely resummed n-rung ladder
 and
studies have reported  a strong  convergence of the BFKL
 series considering
few orders in the expansion (for example, in
 heavy vector meson production
\cite{Ryskin}). Furthermore, there
 is important evidence that the asymptotic
solution to the BFKL
 equation is inapropriated in the most of the HERA range
\cite{Ryskin}, and the
 expansion order by order allows to identify the onset
of the
 region where the full BFKL series resummation is required.

In our previous work on this subject \cite{Trunkep}, we summed
up two terms of the perturbative expansion and obtained the
imaginary part of the DIS amplitude, hence the inclusive
structure function $F_2(x,Q^2)$. Using the most recent HERA data
on the $F_2$-logarithmic slope to determine the adjustable
parameters at the small $x$ region, we performed a broadly
description of the structure function and its slope in the
kinematical range of momentum fraction $x<10^{-2}$. The resulting $F_2$ and
its gluon content turn out having a mild logarithmic growth as $x$ diminishes
in contrast with the steeper increasing from the usual DGLAP or BFKL dynamics,
which we discuss a little more in the next section. It was also
 determined
that the non-perturbative background is not negligible
 in all the  kinematical
region and should be better addressed. A remaining question
 in those previous
works  is the role played by the next order in the
 perturbative expansion and
the modeling of the soft background, object of this work. 
 To address the
issues above, here one performs a more detailed
 study on the inclusive
structure function considering the additional two
 rung contribution for the
truncated series. With the introduction of this new
 term  the virtual
photon-proton cross section reaches the
  behavior settled by the Froissart
bound, i.e.
 $\sigma_{tot}^{\gamma^* p} \sim \ln^2 (1/x)$, however keeping in
mind that it is a result coming from a non-asymptotic series. Moreover,
generally speaking
 the BFKL Pomeron is only a perturbative approximation to
the true Pomeron
 (valid at a limited kinematical range), of which the exact
properties are
 unknown making it necessary to include some contribution of
non-perturbative physics. We included the non-perturbative
 Pomeron in two
quite different forms, described latter on. As a result, the experimental
data
 on the structure function $F_{2}$ are successfully described in a
broader region of $(Q^{2},x)$ variables and a more refined determination of
the
 non-perturbative background is performed.
 
 The final picture  is quite
similar to the two-pomeron one, where the hard and soft pomerons play an
equally important role in data description. 

This paper is organized as
follows. In the next section one introduces the
 main formulae and the most
important features for the inclusive structure
 function, including the two
rung contribution. In the Sec. (3), an
 overall fit to the recent deep
inelastic data is performed based on the
 present calculations and a suitable
modeling for the non-perturbative
 background in two forms are introduced. As
a byproduct, the  resulting slopes are also  shown. Finally, in the last
section we draw our conclusions.
 
 \section{Finite sum of gluon ladders in
deep inelastic scattering}
 
 Regarding the deep inelastic scattering
reaction, the total cross
 section for the process $\gamma^* \,p \rightarrow
X$, where $X$ states for all
 possible final states, is  obtained from Optical
theorem through the imaginary
 part of the elastic $\gamma^*\,p \rightarrow 
\gamma^*\,p$ amplitude. In the
 limit of very high energies the BFKL approach
\cite{LOBFKL} is the most natural
 approach to treat such a process and is
considered to compute the
 correspondent  cross section \cite{Forshaw}. As
already discussed in the
 introduction, the main trouble with that framework
are the unitarity violation
 coming from the LLA Pomeron exchange and the
known strong NLO corrections. In previous calculations we have
 proposed the
truncation of the complete series in order to obtain an
 amplitude described
by a finite sum of gluon ladders \cite{Trunkep}. Below, one
 presents the main
kinematical variables, the calculation of the amplitude up
 to the second
order in the perturbative expansion and the correspondent
 $F_2(x,Q^2)$
expression.
 
Here, we are interested in the high energy region $W^2 >> Q^2$,
where $W$ is the center of mass energy of the system virtual
photon-proton and $Q^2$ is the virtuality of the probe photon.
Defining the momentum fraction, i.e. Bjorken variable, as $x
\approx  \frac{Q^2}{Q^2+ W^2}$, the inequality above implies $x <<
1$, setting  the small $x$ regime. The proton inclusive structure
function, written in terms of the cross sections for the
scattering of transverse or longitudinal polarized photons, reads
as
\begin{eqnarray} F_2(x,Q^2) &=& \frac{Q^2}{4\pi^2 \alpha_{\rm{em}}}
\left[\sigma_T(x,Q^2) + \sigma_L(x,Q^2)\right]. \label{eq1}
\end{eqnarray}

In the asymptotic high energy limit, for photons with
polarization $\lambda$, the cross section is  given by the convolution of a
perturbative kernel, which provides the dynamics of the process, with the
corresponding impact factors of the interacting particles,
\begin{eqnarray}
\sigma_{\lambda}(x,Q^2)=\frac{{\cal G}}{(2\pi)^4} \, \int
\frac{d^2\kk}{\kk^2}\,\frac{d^2\kg}{\kg^2}\,\Phi^{\gamma^*}_{\lambda}(\kk)\,
F(x,\kk,\kg)\,\Phi_p(\kg). \label{eq2} \end{eqnarray}

Clarifying the notation, ${\cal G}$ is the color factor for the
color singlet exchange and $\kk$ and $\kg$ are the transverse
momenta of the exchanged reggeized gluons in the $t$-channel. The
$\Phi^{\gamma^*}_{\lambda}(\kk)$ is the virtual photon impact
factor (with $\lambda = T,\,L$) and $\Phi_p(\kg)$ is the proton
impact factor. The first one is well known in perturbation theory
at leading order \cite{Forshaw}, while the latter is modeled
since in the proton vertex there is no hard scale to allow
pQCD calculations.

The kernel $F(x,\kk,\kg)$ contains the dynamics of the process and
has been systematically determined in perturbative QCD
\cite{Salam}.  The main properties of the LO kernel are well
known \cite{LOBFKL} and the results coming from the NLO
calculations indicate that the perturbative Pomeron can acquire
very significant subleading corrections \cite{Salam}. The most
important feature of the LLA BFKL Pomeron is the leading
eigenvalue of the kernel, leading  to a steep rise with decreasing
$x$, $F(x)\sim \frac{x^{-\varepsilon}}{\sqrt{\ln 1/x}}$, where
$\varepsilon=4\, \overline{\alpha}_s \ln 2 \approx 0.5$.
Therefore,  the inclusive structure function will present a
similar growth at low $x$. Hence, the resulting amplitude and
consequently the total cross section or structure function, at first glance,
overestimates the growing of the observable currently measured.

We proposed an alternative phenomenological way to calculate the
observables at current energies, showing that a reliable
description of both proton-(anti)proton and virtual photon-proton
collisions is obtained considering the truncation up to two
orders in the perturbative expansion \cite{Trunkpp,Trunkep}. In the
accelerators domain
 the asymptotic regime is not reached and there is no
room in
 rapidity to enable an infinite n-gluon cascade, represented
diagrammatically by the BFKL ladder. Moreover, a steep convergence
 of the LO
BFKL series in few orders in the perturbative
 expansion  has been already
reported \cite{Ryskin} and
 phenomenological studies indicate that such a
procedure is
 reasonable at least in proton-proton collision
\cite{Fiore,hepsasha}. In order to perform the calculations,
 it 
should be taken into account the convolution between the photon
and proton impact factors and the corresponding gluon ladder
exchanges in each order. The Born contribution comes from the
bare two gluon exchange and  in the leading  order  the
amplitude is imaginary  at high energies and  written for
$t=0$ as
\begin{eqnarray}
{\cal A}^{Born}(W,t=0)=\frac{2\,\alpha_s \, W^2}{\pi^2}\,\sum_f e^2_f\,
\int \frac{d^2\kk}{\kk^4}\,\Phi^{\gamma^*}_{\bot} (\kk)\,\Phi_p(\kk).
\label{eq3}
\end{eqnarray}

The next order contribution is obtained from the graphs
considering the one rung gluon ladder and has the following
expression in LLA

\begin{eqnarray*} {\cal A}^{\rm{one\!-\!rung}}(W,t=0)= \frac{6 \alpha^2_s
W^2}{8 \pi^4} \sum_f
 e^2_f \ln(W^2/W^2_0) \int \frac{d^2\kk}{\kk^4}
\frac{d^2\kg}{\kg^4}
 \Phi^{\gamma^*}_{\bot} (\kk) K(\kk,\kg) \Phi_p(\kg) \,.
\end{eqnarray*}

The $\alpha_s$ is the strong coupling constant, considered fixed since we
are in the framework of the LO BFKL approach. As a remark, the running of the
coupling constant contributes significantly for the NLO BFKL, since it is
determined by subleading one-loop corrections, for example the self-energy and
vertex-correction diagrams \cite{Salam}. The typical energy of the process is
denoted by $W_0$, which scales the logarithm of energy and takes an arbitrary
value in LLA.

The perturbative kernel  $K(\kk,\kg)$ can be calculated order by order in the
perturbative expansion and is encoded by the BFKL kernel if one considers the
LLA resummation. In the  present case,  $t=0$  and it describes the gluon
ladder
 evolution in the LLA of $\ln (s)$ as already discussed above. The
Pomeron is
 attached to the off-shell incoming photon through the quark loop
diagrams,
 where the Reggeized gluons are attached to the same and to
different quarks in
 the loop  \cite{Trunkep}. Since the
transverse
 contribution dominates over the longitudinal one, hereafter
$\Phi^{\gamma^*}_{\bot}$ is the virtual photon impact factor averaged over the
transverse polarizations \cite{Balitsky},
\begin{eqnarray} \Phi^{\gamma^*}_{\bot}(\kk)=\frac{1}{2} \int_0^1
\frac{d\tau}{2\pi} \,\int_0^1 \frac{d \rho}{2\pi}\, \frac{\kk^2(1-2\tau
\tau^{\prime})(1-2\,\rho \rho^{\prime})}{\kk^2 \,\rho \rho^{\prime} + Q^2
\rho \,\tau \tau^{\prime}}, \label{eq4}
\end{eqnarray}
where $\rho$, $\tau$ are the Sudakov variables associated
to the momenta in the photon vertex and the notation $\tau^{\prime}=(1-\tau)$
and $\rho^{\prime}=(1-\rho)$ is used, Ref. \cite{Forshaw}.

A well known fact is that we are unable to compute the proton
impact factor $\Phi_p(\kg)$ using perturbation theory since it is
determined by the large-scale nucleon dynamics. However,
 gauge invariance requires that  $\Phi_p(\kg \rightarrow
0)\rightarrow 0$ and then the proton impact factor can be
modeled as a phenomenological input obeying that limit and takes a
simple form
\begin{eqnarray} \Phi_p(\kg)={\cal N}_p
\,\frac{\kg}{\kg + \mu^2}, \label{eq5}
\end{eqnarray}
where ${\cal N}_p$ is the unknown normalization of the proton
impact factor and $\mu^2$ is a scale which is typical of the
non-perturbative dynamics. Furthermore, these non-perturbative
parameters can absorb possible subleading contribution in each
order of the perturbative expansion \cite{Trunkep}.

Considering the electroproduction process,
summing the two first orders in perturbation theory  we can write the
expression for the inclusive structure function, whose
contributions have been already discussed (section V of Ref. \cite{Balitsky} )
\begin{eqnarray} F_2(x,Q^2)=
\frac{8}{3}\,\frac{\alpha^2_s}{\pi^2} \sum_f e^2_f \, {\cal N}_p
\left(F_2^{\rm{Born}}(Q^2,\mu^2) + \frac{3\,\alpha_s}{\pi}\,\ln
\left(\frac{x_0}{x}\right) \,F_2^{(\rm{I})}(Q^2,\mu^2)\right)\, \label{eq6}
\end{eqnarray}
where the functions $F_2^{\rm{Born}}(Q^2,\mu^2)$ and
$F_2^{(\rm{I})}(Q^2,\mu^2)$ ($\rm{I}$ meaning one-rung contribution) 
given by

\begin{eqnarray}
\label{eq7} & & F_2^{\rm{Born}}(Q^2,\mu^2) = \frac{1}{2} \ln^2 \left(
\frac{Q^2}{\mu^2} \right) + \frac{7}{6}\ln \left( \frac{Q^2}{\mu^2}
\right) + \frac{77}{18} \,,\\ & & F_2^{(\rm{I})}(Q^2,\mu^2) = \frac{1}{6} \ln^3
\left( \frac{Q^2}{\mu^2} \right) + \frac{7}{12}\ln^2 \left(
\frac{Q^2}{\mu^2} \right) + \frac{77}{18}\ln \left( \frac{Q^2}{\mu^2}
\right) + \frac{131}{27} + 2\,\zeta (3)\, \nonumber
\end{eqnarray}
where  $x_0$ gives the scale to define the logarithms on energy for the
LLA BFKL approach (is arbitrary and enters as an additional parameter) and
$\zeta
 (3)=\sum_r(1/r^3) \approx 1.202$ is the Riemann $\zeta$-function.

Connecting
the present result to the Regge phenomenology, the truncation of the
perturbative series reproduces the main characteristics coming
from  the Regge-Dipole Pomeron model.  In the Dipole Pomeron the
structure function is  $F_2(x,Q^2)\sim R(Q^2)\ln (1/x)$, whose
 behavior  corresponds to the contribution of a double $j$-pole to the
partial amplitude of $\gamma^* p \rightarrow \gamma^* p$, where $R(Q^2)$ is
the Pomeron residue function.  Moreover, the
Dipole Pomeron  trajectory has unit intercept, $\alpha_{\pom}(0)=1$, and has
been used in several phenomenological fits to hadronic sector and HERA data
\cite{dippom}.   While in the Dipole Pomeron picture the
residue is factorized from the energy behavior in the amplitude,
the $Q^2$ dependence is calculated order by order in perturbation
theory  in our case. 

In Ref. \cite{Trunkep}, we choose to determine the parameters of the model
($\mu$, $x_0$ and ${\cal N}_p$) from a smaller
 data set, meaning  the
latest HERA measurements on the $Q^2$-logarithmic
 derivative reported by the
H1 \cite{H1c1} and ZEUS Collaboration (preliminary).
 The reasons for that are
the consistent and precise measurements of the slope and the additional fact
that using the $Q^2$ derivative we avoid contributions from the
non-perturbative background depending weakly on $Q^2$. The obtained values
were  consistent with the naive estimates, namely $\mu^2$
 would be  in the
non-perturbative domain ($\mu \approx \Lambda_{QCD}$) in
 both data sets and
$x_0$ has a value consistent with the Regge limit. The
 fitted expression to
the slope described the data  with good agreement,
 producing effectively the
same linear behavior in $\ln Q^2$ considered by the
 H1 Collaboration fitting
analyzes. Considering the $x$ dependence of the
 slope, a gluon
distribution softer than those coming from the
 usual approaches, $xG(x,Q^2)
\sim x^{-\lambda}$ \cite{DGLAP,LOBFKL} is obtained towards small
 $x$.
  The growth
of the structure function  shown
 large deviations from the steep increasing
present at both  LO BFKL series
 and DGLAP approach, where $F_2 \sim
x^{-\lambda}$. The non-perturbative
 contribution (from the soft dynamics),
mainly at low $Q^2$ virtualities,  was
 found  not
negligible. One estimated that those effects
 imply correction of
$\approx 20$ \% in the overall normalization.

In the next section we
address the effects of introducing the
 next order in the perturbative
expansion and a more refined parametrization for
 the non-perturbative
background. In  order to do so, we need to calculate the two-rung
contribution in the ladders summation. Following the calculations
\cite{Trunkep,Balitsky}, it results
 
\begin{eqnarray}
 \label{eq8}
 F^{(\rm{II})}_2(x,Q^2) & = &
\frac{8}{3}\,\frac{\alpha^2_s}{\pi^2} \sum_f e^2_f \, {\cal N}_p \, \left[
{1\over 2} \left({3\alpha_s\over \pi}\ln {x_0\over x}\right)^2
 \left({1\over
24}\ln^4{Q^2\over \mu^2}+{7\over 36}\ln^3{Q^2\over
 \mu^2}+  {77\over
36}\ln^2{Q^2\over \mu^2 } \, + \right. \right.  \nonumber \\
  & &  + \,\,
\left. \left. ({131\over 27}+4\zeta(3))  \ln{Q^2\over \mu^2}+
  {1396\over
81}-{\pi^4\over 15} +{14\over 3}\zeta(3)
   \right) \right],
\end{eqnarray}
where the notation is the same one of the expressions above and such a
contribution should be added to the  previous results in order to perform a
fit to the
 HERA data.  Our final fitting expression is then given by
\begin{eqnarray}
F_2(x,Q^{2})=F_{2}^{\rm{Born}}  + F_{2}^{(\rm{I})}\,[\rm{one\!-\! rung}]
+F_{2}^{(II)}\,[\rm{two\!-\! rung}]+  F_2^{\rm{soft}}\,[\rm{Background}].
\label{fittexp}
\end{eqnarray}

Some remarks about the results derived from the truncation of the
perturbative series are in order. An  important issue emerging is
the role of the  subleading contributions at a  fixed order
expansion of the BFKL approach,  firstly
addressed by Fiore et al. in the hadronic sector \cite{Fiore}.
The coefficients of different powers of $\ln(s)$ in the series
refer to the dominant contribution, at asymptotic energies, for
each perturbative order. However,  performing a
phenomenological description of data the subleading contributions
are absorbed into the normalization constants for each ladder in
the expansion, or in the non-perturbative parameters of the model. This can
shed light about the strenght of those contributions in the description of the
 observables. For instance, in Ref. \cite{Fiore} it was found that already at
the Tevatron energy range, the coefficient weighting the term $\ln^3(1/x)$ is
approximately  equal to zero in contrast with the expectations from the
perturbative coefficient calculated in the complete series.
 This is the main
feature of reliability  in the data
 description of Refs.
\cite{Fiore,hepsasha} and in our more
 detailed study for $pp$ $(p\bar{p})$
reactions in Ref.
 \cite{Trunkep}.

Concerning the gluon content of $F_2$, in general grounds the truncated series
provides a mild logarithmic growth  instead a steep one from DGLAP or LO BFKL
resummations. In order to illustrate the result coming from adding rungs in
the ladder, that is summing new terms in the truncation, we can make use of
the effective exponent of $F_2$. One can parametrize the
increasing of the structure function (and the gluon content at small-$x$) in
the simple form $F_2(x, Q^2)\simeq xG(x,Q^2)\approx
f(Q^2)\, x^{-\lambda_{eff}}$, where we can calculate the exponent as $
\lambda_{eff}= d \ln F_2/d\ln (1/x)$. 

The Born two gluon approximation to $F_2$, i.e. first term in Eq. (6),
provides a constant contribution on $\ln (1/x)$ and therefore it gives
$\lambda_{eff}=0$. Of coarse, this result is rule out by the experimental
measurements. When one introduces the one-rung piece,  turn out
$F_2=f_0(Q^2) + f_1(Q^2)\ln (1/x)$, where $f_0$ and $f_1$ are the
$Q^2$-dependent (logarithmic) factors appearing in Eq. (6). The resulting
effective exponent is thus $\lambda_{eff}=[f_0/f_1 + \ln (1/x)]^{-1}$. If we
consider the approximation $f_0/f_1\ll 1$ at fixed $Q^2$, which it  seems
reasonable from inspection of  Eqs. (7,8), one estimates that the effective
intercept takes values between $\lambda_{eff}\simeq 0.2$ at $x=10^{-2}$ and
decreasing to $\lambda_{eff}\simeq 0.1$ for $x\sim 10^{-5}$. This growth is
softer than the typical DGLAP result $\lambda_{eff}\simeq 0.3$. Concerning the
addition of the two-rung piece, the result is less simple; however if we
consider the limit case where $f_2(Q^2) \gg f_1(Q^2), f_2(Q^2)$, one obtains 
$\lambda_{eff}\approx 2/\ln (1/x)$. Then, the effective exponent would take
values $0.4$ at $x=10^{-2}$ and saturating at  $0.2$ around $x\sim 10^{-5}$:
values quite similar to the mean value from DGLAP expectations. The results
estimated above have implications for instance on diffraction. The $x_{\pom}$
dependence of the diffractive structure function, $F_2^D$, depends directly on
the unintegrated gluon distribution ${\cal F}(x_{\pom}, k_T)=k_T^2\,\partial\,
xG(x,k_T^2)/\partial \,k_T^2$ and then one expects a mild logarithmic
increasing and a non-trivial (logarithmic) $k_T$ dependence.

Now, having the expression for the inclusive structure
function at hand, in the next
 section we compare it with the HERA
experimental results, determining the
 adjustable parameters and the range of
validity for this model.
 
\section{Fits to the HERA data: choice of background and discussion of the
results.}
 
In order to compare the expression obtained to  the structure function
$F_2(x,Q^2)$, Eq. (\ref{fittexp}), with the experiment, we choose to use the
updated
 HERA data set starting from the smallest available $Q^2$ for  small
$x$ region $x \approx 10^{-2}$. The BFKL Pomeron as well as
 its truncation
should be valid in a specific kinematical interval,
 for instance the full
series \cite{fullseries} can accommodate
 data ranging from $ 1 \leq Q^2 \leq
150 $ GeV$^2$ in a good
 confidence level. For this purpose we need to include
a soft
 background, accounting for the non-perturbative content and 
providing a smooth transition to $Q^{2}=0$. Notice that an
extrapolation to the photoproduction region  is still lacking in
the present analysis, although we have a definite expression for
that \cite{Trunkep}. Thus, here we consider the
electroproduction case and the low virtualities range will be
driven by the non-perturbative background, modeled through a soft
Pomeron.   There exists numerous possibilities from novel
\cite{DL,Cudell,DM} to previous \cite{CKMT} models, which   give significant contribution in  the data
description for $Q^2 \leq 10$ GeV$^2$. In our case, we  use the model with
the most economical  number of
 parameters and for  this purpose  we have
first  selected the latest
 version \cite{KMP} of the CKMT model \cite{CKMT}:
\begin{equation} \label{eq9} F^{\rm{soft}}_2(x,Q^2)= A\left(
\frac{x_0}{x} \right)^{\Delta(Q^2)} \left(
\frac{Q^2}{Q^2+a}\right)^{\Delta(Q^2)+1}\,,  \end{equation} where
the expression for $\Delta(Q^2)$ has the following form:
\begin{equation} \label{eq10} {\Delta(Q^2)}= \Delta_{0} \left( 1+
\frac{Q^2 \Delta_{1}}{\Delta_{2}+Q^2}\right). \end{equation}
where $\Delta(Q^2)$ is the Pomeron intercept and the remaining parameters
are defined in \cite{CKMT}.  Such a model considers a soft Pomeron which is
a single pole in the complex
 angular momentum plane having a $Q^2$-dependent
intercept. Although formally
 it is not a pure Regge approach, it describes
the low virtuality region with very
 good agreement \cite{CKMT} with a limited
number of adjustable parameters. The dependence
 on $Q^2$ of the structure
function comes from the Pomeron residue and in
 general is modeled  since
there is
 little theoretical knowledge, namely  vertex functions and couplings
at the amplitudes. The gauge invariance only requires that it should
vanish as $Q^2 \rightarrow 0$. Particularly, a model-independent way is more 
preferable, for instance as
 performed by \cite{DL}, where the residue
function is extracted from data and then fitted with a suitable adjusting
function.

 Another possibility  is to
 select a Pomeron for the
background which has an intercept equal
 to $1$ and has the form of a
non-perturbative truncated $\log
 (\frac{Q^{2}}{x})$ series (soft multipole
Pomeron) \cite{Cudell,DM}, with the form  \begin{equation}
\label{eq11} F^{\rm{soft}}_{2}(x,Q^{2})=Q^{2} \left[ a \cdot \left(
\frac{d}{Q^{2}+d} \right)^{\alpha} + b\cdot \ln \left( \frac{Q^{2}}{x}
\right) \left( \frac{d}{Q^{2}+d} \right)^{\beta} + c\cdot \ln^{2} \left(
\frac{Q^{2}}{x} \right) \left( \frac{d}{Q^{2}+d} \right)^{\gamma} \right]\,,
\end{equation}

where this choice has a larger number
of parameters  than the previous background and we naively expect a better 
accommodation to the  data. 
To proceed, we used two models for the background to fit the
structure function  $F_{2}$. The finite sum of gluon ladders is
encoded in the Eqs. (\ref{eq6}) and (\ref{eq8}), while the soft
Pomeron is given by Eqs.  (\ref{eq9}) and (\ref{eq11}). In both
cases we have applied the factor $(1-x)^7$ to provide the behavior of
$F_{2}$ at large $x-$region and being the same for both soft and
hard Pomeron, extending  the applicability of the models to the $x\rightarrow
1$
 region. From the dimensional-counting rules these threshold correction
factors  are given by
 $(1-x)^{2n-1}$, where $n$ is the spectators number (for
the
 Pomeron it is equal 4). Thus, our considerations are consistent
 with this
fact and should provide a good description even at
 large $x$.

For the fitting procedure we consider the data set containing all
available HERA data for the proton structure function $F_2$ \cite{H1c1},
\cite{ZEUSc1,H1c6}. Notice that
 the most recent measurements in H1 and
ZEUS are more accurate (stat. error
 $\approx 1$ \%) than the previous ones,
providing stringest constraints to
 the parameters. For the fit we have used
$496$ experimental points for
 $x\leq0.025$ and $0.045\leq Q^2 \leq 1500$
$GeV^2$. We selected the
 overall normalization factor as a free parameter for
the hard Pomeron
 contribution, Eq. (\ref{eq6}), defined as ${\cal
N}=\frac{8}{3}
 \frac{\alpha^2_s}{\pi^2} \sum\limits_{f} e^2_f {\cal N}_p$,
considering
 four active flavours. In the Figs. (1)-(3) we show the resulting
fits considering the  two  distinct backgrounds. The best fit parameters of
the  model are shown in Tables (1,
 2).

\begin{center}
\begin{tabular}{|cccccccccc|}
\hline \multicolumn{1}{|c|}{$\cal N$} &
\multicolumn{1}{|c|}{$\mu^2$} & \multicolumn{1}{c|}{$x_0$} &
\multicolumn{1}{c|}{$\alpha_{s}$} & \multicolumn{1}{c|}{$A$} &
\multicolumn{1}{c|}{$a$} & \multicolumn{1}{c|}{$\Delta$} &
\multicolumn{1}{c|}{$\Delta_{1}$} & \multicolumn{1}{c|}{$\Delta_{2}$} &
\multicolumn{1}{|c|}{$\chi^{2}$} \\
\hline \multicolumn{1}{|c|}{$0.00312$} &
\multicolumn{1}{|c|}{$1.39$} & \multicolumn{1}{c|}{$0.251$} &
\multicolumn{1}{c|}{$0.2$ (fixed)} & \multicolumn{1}{c|}{$0.279$}
& \multicolumn{1}{c|}{$0.579$} & \multicolumn{1}{c|}{$0.108$} &
\multicolumn{1}{c|}{$1.65$} &
\multicolumn{1}{c|}{$9.68$} & \multicolumn{1}{|c|}{$1.14$} \\
\hline
\end{tabular}

\smallskip\

{\bf Table 1:} Parameters of the model with the first background, Eq. (9), 
obtained
 from the fit.
\end{center}

\begin{center}
\begin{tabular}{|cccccccccccc|}
\hline \multicolumn{1}{|c|}{$\cal N$} & \multicolumn{1}{|c|}{$\mu^2$} &
\multicolumn{1}{c|}{$x_0$} & \multicolumn{1}{c|}{$\alpha_{s}$} &
\multicolumn{1}{c|}{$a$} & \multicolumn{1}{c|}{$b$} &
\multicolumn{1}{c|}{$c$} & \multicolumn{1}{c|}{$\alpha$} &
\multicolumn{1}{c|}{$\beta$} &
\multicolumn{1}{c|}{$\gamma$} & \multicolumn{1}{|c|}{$d$} & \multicolumn{1}{c|}{$\chi^{2}$} \\
\hline \multicolumn{1}{|c|}{$0.0191$} &
\multicolumn{1}{|c|}{$0.593$} & \multicolumn{1}{c|}{$0.148$} &
\multicolumn{1}{c|}{$0.242$} & \multicolumn{1}{c|}{$0.506$} &
\multicolumn{1}{c|}{$-0.426$} & \multicolumn{1}{c|}{$-0.0495$} &
\multicolumn{1}{c|}{$0.491$} & \multicolumn{1}{c|}{$1.69$} &
\multicolumn{1}{c|}{$0.727$}
& \multicolumn{1}{c|}{$0.130$} & \multicolumn{1}{c|}{$0.94$}\\
\hline
\end{tabular}

\smallskip\

{\bf Table 2:} Parameters of the model with the second background, Eq. (11), 
obtained
 from the fit.
\end{center}

Before to proceed, we discuss how the results presented here compare with the
previous ones using only two terms of the perturbative expansion and how  the
play roled by the background is affected by adding a new term in the
expansion.  Considering only
 the finite sum of gluon ladders (hard Pomeron)
one obtains the
 following:  using either the  one rung ladder [Eq.
(\ref{eq6})] as well as the two-rung contribution
[Eqs.
(\ref{eq6}),(\ref{eq8})], the model provides the same
{$\chi^{2}/\rm{dof}=1.6$} in the interval $1.2\leq Q^2 \leq  150$
 GeV$^2$,
whereas the fit is degraded for a larger interval of
 virtualities (i.e.
$\chi^{2}/\rm{dof}\approx 2.5$ for $1.2\leq
 Q^2 \leq 800$ GeV$^2$). These
results are in agreement with the
 analysis \cite{fullseries} (limited to a
specific H1 data set \cite{H1c4}),
 corroborating that the BFKL-like models
accommodate data at
 virtualities up to $\approx 150$ GeV$^2$ \cite{vicgay}.
This kinematical limit corresponds to the region where DGLAP resummation start
to be important, that is the $[\alpha_s \ln (Q^2/Q_0^2)]^n$ terms would
dominate the $\alpha_s \ln (1/x)]^n$ contributions. The
resulting $\chi^{2}/\rm{dof}=1.6$ is still formally large in comparison with
more phenomenological approaches, however it
 can be justified if we note the
little number of parameters considered (3
 adjustable constants) and
desconsidered the effects of large $x$.
 The main point is that the addition
of a new term does not improve strongly the description of data. This fact is
supported by the finding that the coefficient weighting the term $\ln^2 (1/x)$
in the expansion is significantly small in conparison with the $\ln (1/x)$ term
\cite{Fiore}.

Concerning the role played by the background, the procedure used in the
previous work \cite{Trunkep} was somewhat different. We choose the slope data
for fitting the parameters of the model for two reasons: (a) they are quite
precise and are determined from the updated measurements of $F_2$; (b)
considering a derivative on $Q^2$ we are avoiding the contributions from the
non-perturbative background (depending weakly on $Q^2$), which should be added
to the final $F_2$ expression.  There, we have multiplied the resulting $F_2$ 
by  a factor 1.2, in order  to rougly simulate the background 
effects. Of course, this is only a naive estimation and leads to a worse
description of data. However, the main point  is that the background is
important in data description over all the kinematical range and would be of
order $\approx 20 \%$. Here, we have modeled carefully the soft background
based on Regge phenomenology for the Pomeron. From the discussion  about
the resulting effective exponent in the previous section, we naively expect
that the background in the one-rung case would have a larger effect in mimic
the measured $\lambda_{eff}$ than in the two-rung case.  Although we are not
able to produce a reasonable comparison with the previous work due to the
different procedure considered, the definitive finding is that the background
is active  in the full kinematical interval, instead of being important only
in a specific range, and account for a non-negligible contribution to $F_2$. 

Returning to the present analysis, the final result considering the
additional soft background, is in agreement  with data in a good confidence
level for a large span of $Q^2$. The $\chi^{2}$ values obtained select the
logarithm-type as the better background. The negative
value for the $c$ constant is a non-favoured choice despite
the better $\chi^{2}/\rm{dof}$. Instead,  if we consider only a $\sim
\log(Q^2/x)$ or $\sim \log^2(Q^2/x)$ parametrization one gets a smaller number
of parameters, however the  $\chi^{2}/\rm{dof}$ value becomes similar to the
CKMT-type.

In Fig. (1) are shown the results for the low $Q^2$ data, expected to be 
dominated by the soft Pomeron background. Both choices for the soft Pomeron
seem to describe these data bins quite  well, however the CKMT-type one
provides a
 steeper increasing with $x$ than the log-type up to $\approx 0.25$
GeV$^2$, which  comes directly from the respective intercept for each
background; above that virtuality  
 both have the same behavior on
Bjorken $x$.
 
 \begin{figure}[t]
\begin{center}
\includegraphics*[scale=0.5]{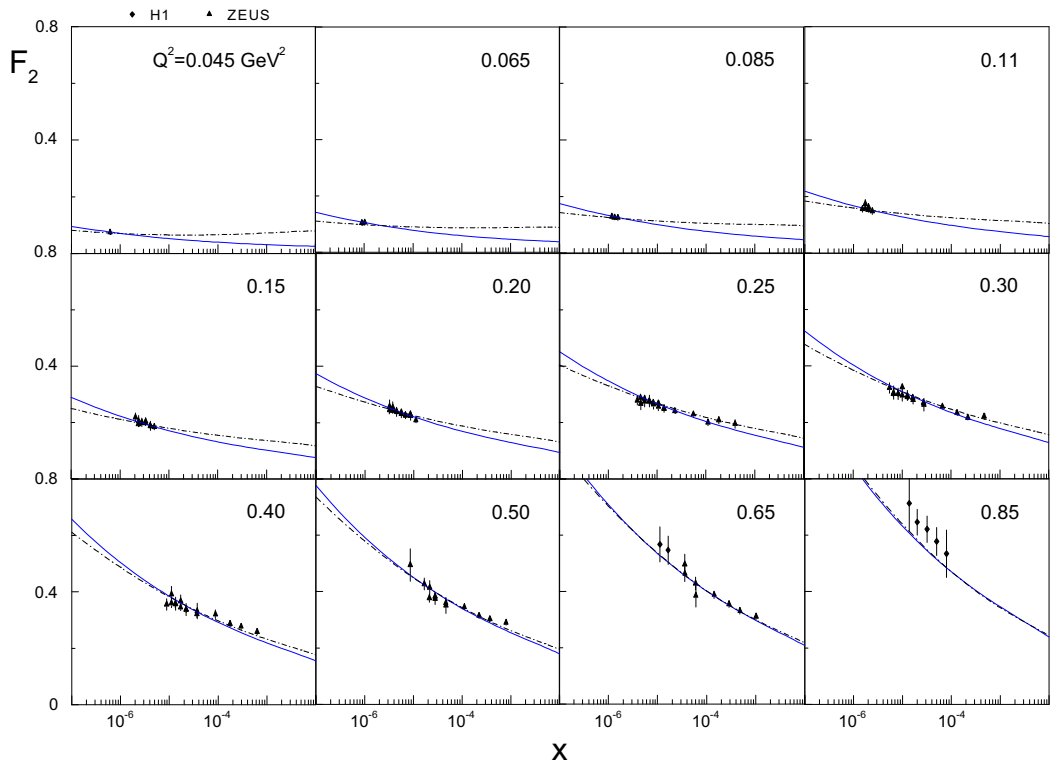}
\caption{\small The  inclusive structure function at
very small $Q^{2}$ virtualities. The solid line corresponds to
 the 
model with the first background [Eq. (10), (11)], while the dash-dotted line to
the model
 with the second background [Eq. (12)].}
\end{center}
\end{figure}

In Figs. (2-3), one verifies that the description is independent
of the specific choice for the soft Pomeron, as  expected,
since in that kinematical region the finite sum of ladders dominates. At
higher $Q^2$ the description deviates
 following the different backgrounds:
at
 $Q^2\approx 90$ GeV$^2$ and increasing as virtualities get larger.
However, the deviation is present in a kinematical region where
 no data is
measured and does not allow an unambiguous conclusion.
 In general grounds, as
it is seen from the figures and tables the
 model of  finite sum of ladders
 with the logarithmic-type background
 describes better the entire data set 
and probably it would be even better
 for a wider interval of $x$ and
$Q^{2}$, but the CKMT-type
 contribution is phenomenologically preferred due to
the smaller
 set of parameters considered.
 
 \newpage

\begin{figure}[p]
\begin{center}
\includegraphics*[scale=0.75]{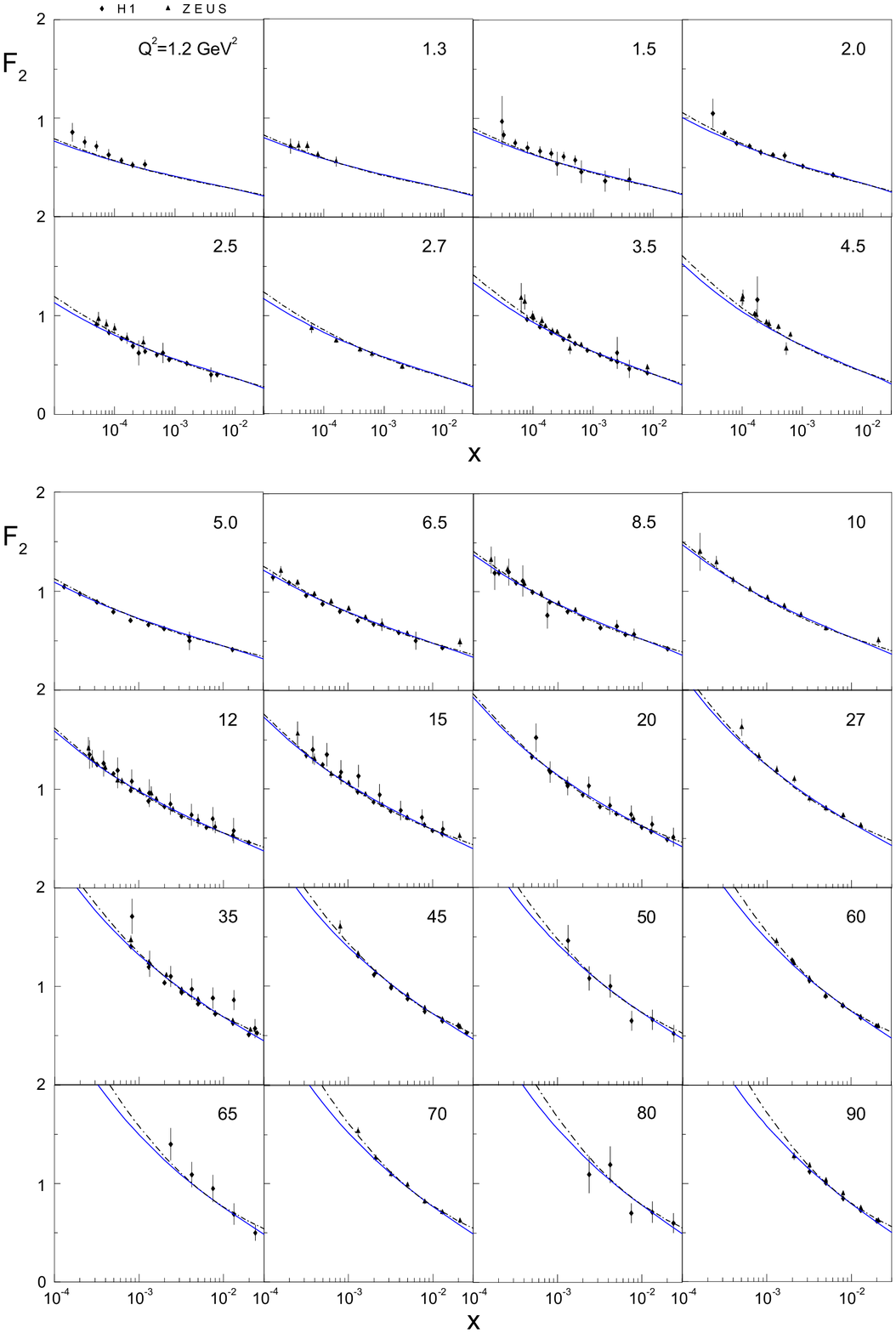}
\caption{\small The  inclusive structure function at
small and medium $Q^{2}$ virtualities. Same notation of the previous
plot.}
 \end{center}
\end{figure}

\begin{figure}[p]
\begin{center}
\includegraphics*[scale=0.35]{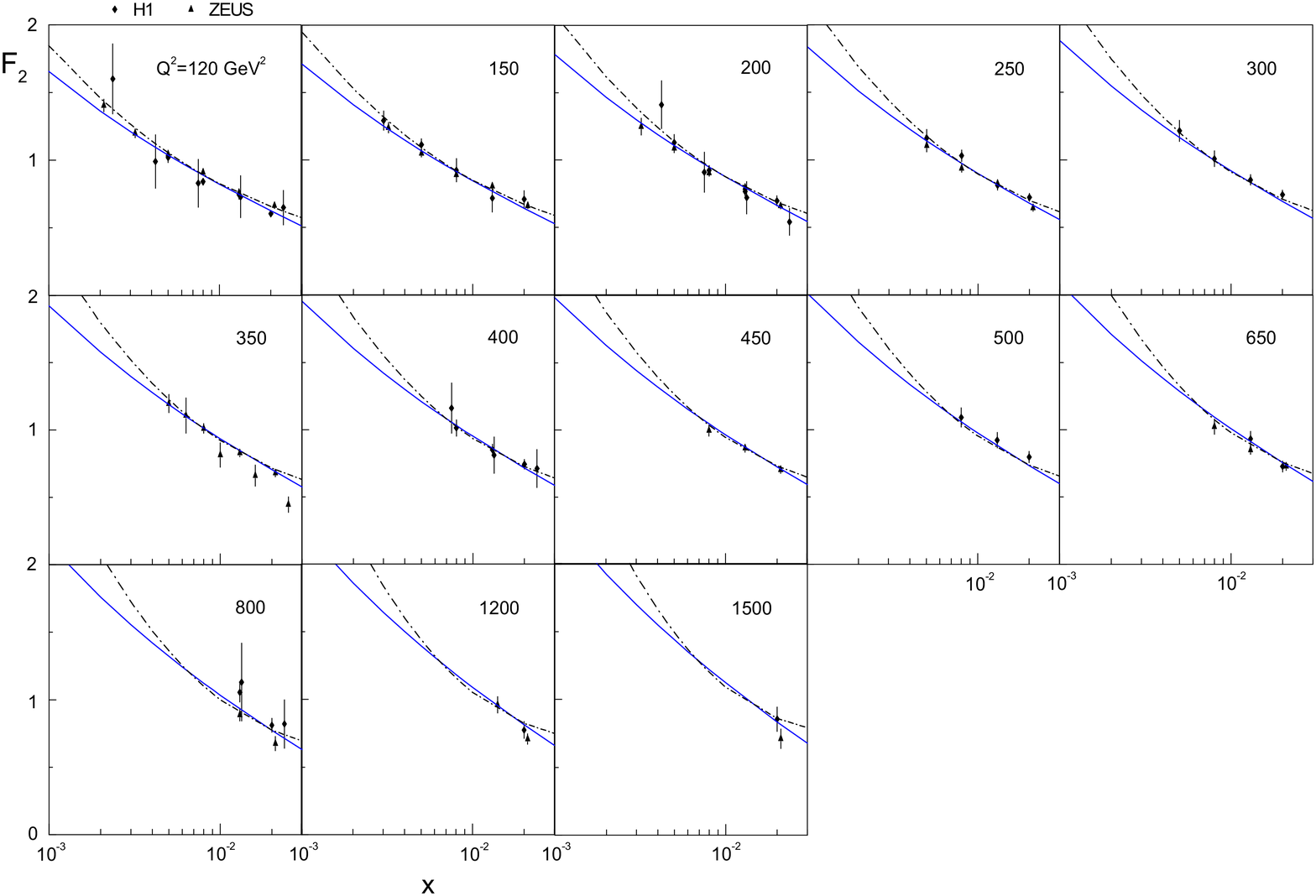}
\caption{\small The  inclusive structure function at
large $Q^{2}$ virtualities. Same notation of the previous plots.}
\end{center}
\end{figure}

Comparing the present analysis with the available Regge phenomenology, we have
a description similar to the two-Pomeron approach  \cite{DL}: the hard
Pomeron in our case is given by the finite sum of gluon ladders up to the
two-rung contribution and the  $Q^2$-dependence is completely determined
from the
 perturbative expansion, truncated at order $\alpha_s^4$. An
extrapolation to
 the photoproduction region is still lacking, which would be
obtained once the
 impact factor of the photon is provided at $Q^2=0$ and
possible singularities
 are regulated. The hard Pomeron couples to each quarks
flavour  with the same coupling, i.e. it is flavour blind. In a similar way as
\cite{DL}, we naively expect roughly to  describe the charm content
$F_2^{c\bar{c}}$ taking $\frac{2}{5}$ of the hard-pomeron contribution (the
fraction means $e_c^2/(e_u^2+e_d^2+ e_s^2)$). We verify that the  backgrounds
play a significant role not only at
 small $Q^{2}$, but in the whole interval
considered, while the pieces
 separately give  good results just  for a
relatively narrow $x$-region.
 Undoubtedly this hybrid approach is good for
all $x$ and $Q^{2}$ if we use a
 proper behavior of $F_{2}$ for large $x$,
i.e. the use of an additional
 non-singlet term.

The model is still comparable with other Regge models, for instance
considering the very nice comparative phenomenological analysis of Ref.
\cite{DM}. There, the two pomeron model,  the soft dipole
Pomeron, a modified two pomeron model and a generalized logarithmic
Pomeron are analyzed  in a large span of $Q^2$ (including photoproduction). At
small-$x$, our data description is in agreement with them, having a similar
$\chi^{2}/\rm{dof}$ (slightly larger for the CKMT background) and with a
similar number of free-parameters (even smaller, for instance using CKMT
background).  The same conclusions are valid concerning the logarithmic
fitting of $F_2$ taken into account in Ref. \cite{Cudell} and in the analysis
of Ref. \cite{DLM}. The main advantage of the present approach is that the
dependence on $Q^2$ of the hard pomeron is completely determined from pQCD, in
contrast with a model-dependent parametrization of the residue pomeron
function in the Regge case.

Considering the QCD fit using the full BFKL series \cite{fullseries}, our
results without a background presents a slightly larger $\chi^{2}/\rm{dof}$ in
a same fitting region ($1.5 \leq Q^2 \leq 150$ GeV$^2$). However, the good data
description coming from \cite{fullseries} is obtained to the cost of a quite
low value of the coupling constant (leading a lower intercept for the BFKL
pomeron), maybe it mimics NLO corrections to the pomeron intercept. In our
case, we have fixed the coupling constant in the reasonable value
$\alpha_s=0.2$, consistent with the virtualities where the fit is performed, 
and have noticied that the effect of left it free is almost negligible, for
instance when we consider the logarithmic background. 

In order to perform a
detailed comparison between the results above  we
 calculated the slopes of
the proton structure function, which are presented in
 the next section.

\section{ The Logarithmic
derivatives}  
The slopes of the proton structure function give valuable
information  concerning the behavior of the gluon distribution and the
effective  Pomeron intercept.  A considerably broad region of DIS kinematical
region in  the upcoming accelerator, i.e. THERA, will effectively enables
to  probe the  saturation phenomenon and other asymptotic properties,
which can be more explicit in derivative quantities directly dependent
on the gluon content of the proton. For this study it is most 
convenient to consider the logarithmic slopes, defined as follows   
\begin{equation}  \label{12} B_{Q}(x,Q^{2})=\frac{\partial
F_{2}(x,Q^{2})}{\partial \ln  Q^{2}},  \end{equation} 
and 
\begin{equation} 
\label{13} B_{x}(x,Q^{2})=\frac{\partial \ln F_{2}(x,Q^{2})}{\partial \ln 
(1/x)}. 
\end{equation} 
 
For numerical calculations of the slopes we used the following 
expressions (see \cite{KMP}), 
\begin{equation} 
\label{14} B_{Q}(x,Q^{2})=\frac{Q^{2}}{2\Delta Q^{2}} \left[ F_{2} \left( 
x,Q^{2}+\Delta Q^{2}\right) - F_{2} \left( x,Q^{2} - \Delta 
Q^{2}\right)\right], 
\end{equation} 
\begin{equation} 
\label{15} B_{x}(x,Q^{2})=-\frac{x}{2\Delta x} \left[ \frac{F_{2} \left( 
x+\Delta x,Q^{2} \right) - F_{2} \left( x - \Delta x,Q^{2} 
\right)}{F_{2}(x,Q^{2})} \right]. 
\end{equation} 
 
The early indirect measurements of the slopes contained the 
shortcoming of correlated bins in the ($x,\,Q^2$) plane 
\cite{ZEUSc3}. The most recent determinations have a better statistics 
and the kinematical variables are no longer correlated 
\cite{H1c1}. Therefore, instead of using the previous two 
types of slopes \cite{ZEUSc3}, which are determined refering two 
variables, and where one is being averaged (one of them is bounded by a
kinematical constraint), we calculated the following   four  types: 
\begin{eqnarray*} B_{x}(x,Q^{2}_{fix}), \quad B_{Q}(x,Q^{2}_{fix}), \\ 
B_{x}(x_{fix},Q^{2}), \quad B_{Q}(x_{fix},Q^{2}). \end{eqnarray*} 
 
Below, we discuss the results for each slope. In the 
Fig. (4) is shown the slope $B_{Q}(x,Q^{2}_{fix})$ in 
virtualities ranging from $2\leq Q^2 \leq 100$ GeV$^2$. The 
calculated slopes are in agreement with data, describing well the 
$x$-dependence. Moreover, the results using the CKMT-type 
background lie slightly above the logarithmic-type at all 
kinematical interval. At larger $Q^2$ the difference between them 
is more evident, however in such a region there are no available  data 
 to provide  a discrimination. 
 
\begin{figure}[ht] 
\begin{center} 
\includegraphics*[scale=0.35]{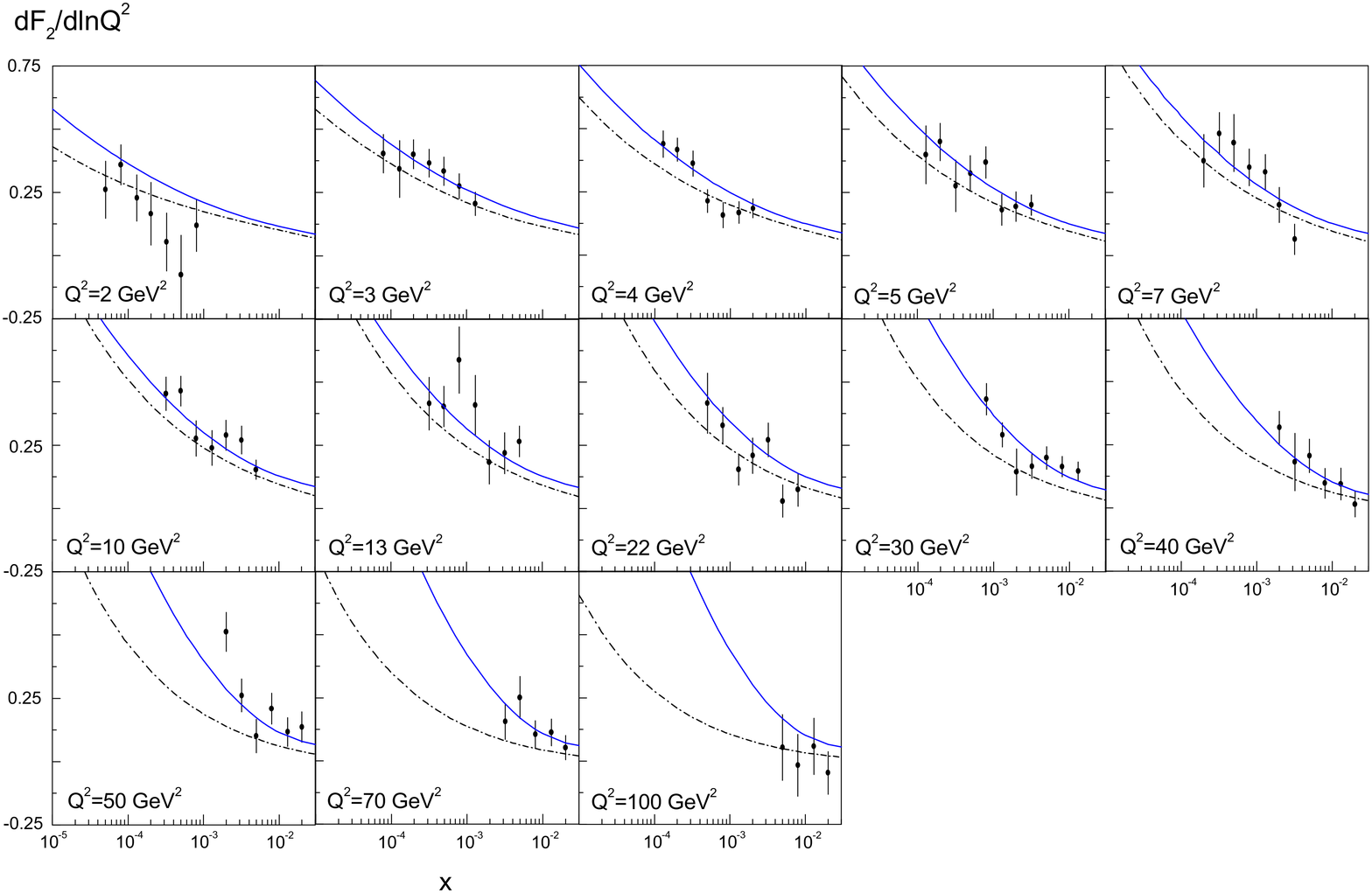} 
\caption{\small The result for the $B_{Q}$ slope plotted as a 
function of $x$ for fixed $Q^{2}$ virtualities compared 
with the latest H1 data \cite{H1c1}. The solid line corresponds to 
model with the first background [Eq. (9)], while the dash-dotted line to the
model  with the second background [Eq. (11)].} 
\end{center} 
\end{figure} 
 
In Fig. (5) are shown the results for the $ B_{Q}(x_{fix},Q^{2})$ for Bjorken 
variable ranging from $8.10^{-5} \leq x \leq 0.02$. In the regions where data 
exist both backgrounds hold. The CKMT-type background select parameters such 
that the description is similar to the results using a QCD fit (H1) 
\cite{H1c1}. The logarithmic-type one presents a typical feature: there is a 
turn-over (a bump) in virtuality $Q^2 \approx 15$ GeV$^2$, in a similar way as 
the Regge dipole Pomeron \cite{DLM}. Both coincide starting at 
$x=3.2\,10^{-3}$. 
 
\begin{figure}[ht] 
\begin{center} 
\includegraphics*[scale=0.35]{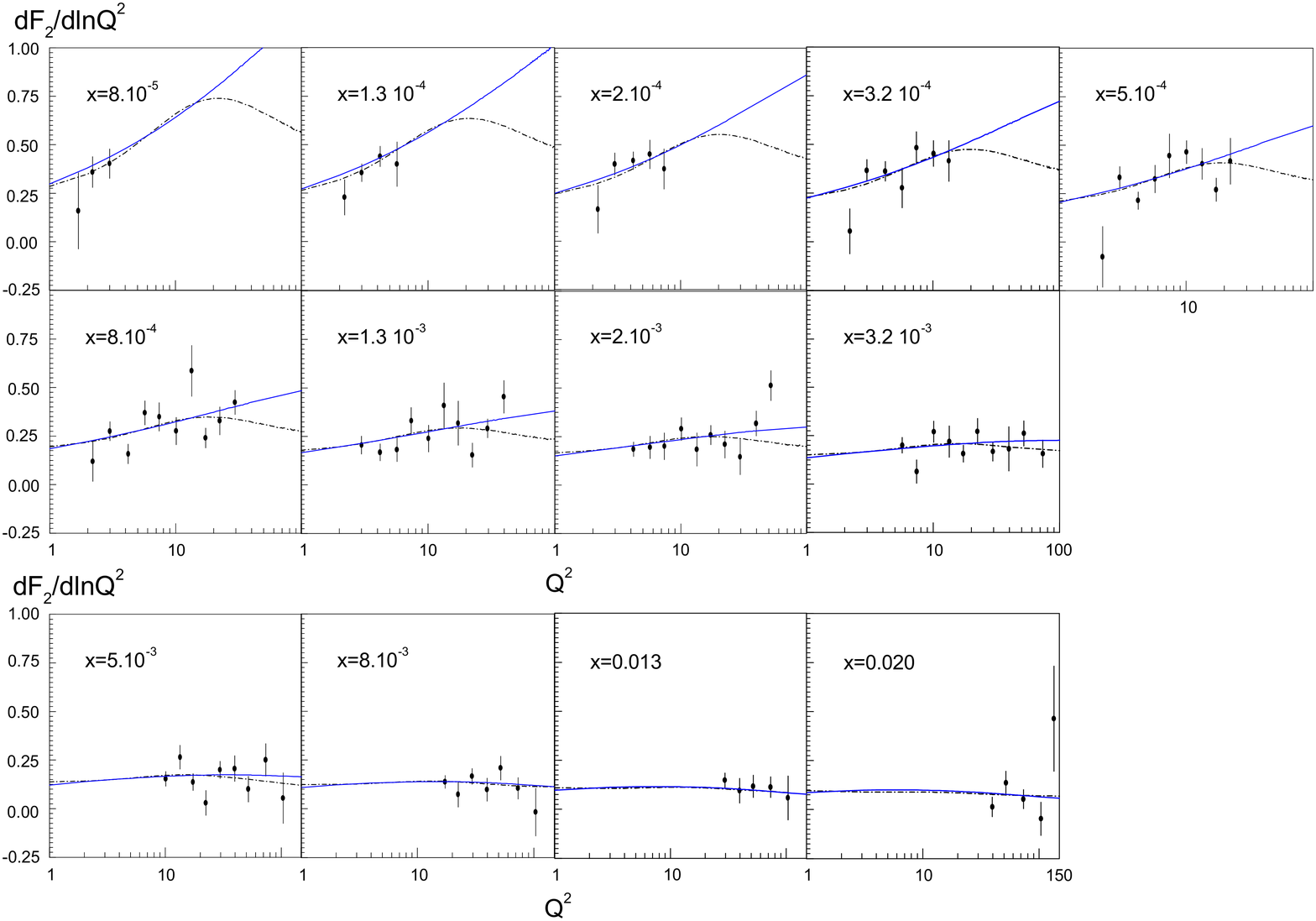} 
\caption{\small The result for the $B_{Q}$ slope plotted as a 
function of $Q^{2}$ (in $GeV^{2}$) for fixed $x$ compared 
with the latest H1 data \cite{H1c1}. Same notation of the previous plots.} 
\end{center} 
\end{figure} 
 
In Fig. (6) are shown the results for the $B_{x}(x_{fix},Q^{2})$ 
for Bjorken variable ranging from $7.10^{-5} \leq x \leq 0.019$. 
Such a quantity corroborates a phenomenological Pomeron having 
$Q^2$-dependent intercept. For this slope, practically the same 
result is obtained considering the two models for the background. 
A significative deviation is verified only at larger $x\sim 0.012$ .
 
\begin{figure}[ht] 
\begin{center} 
\includegraphics*[scale=0.33]{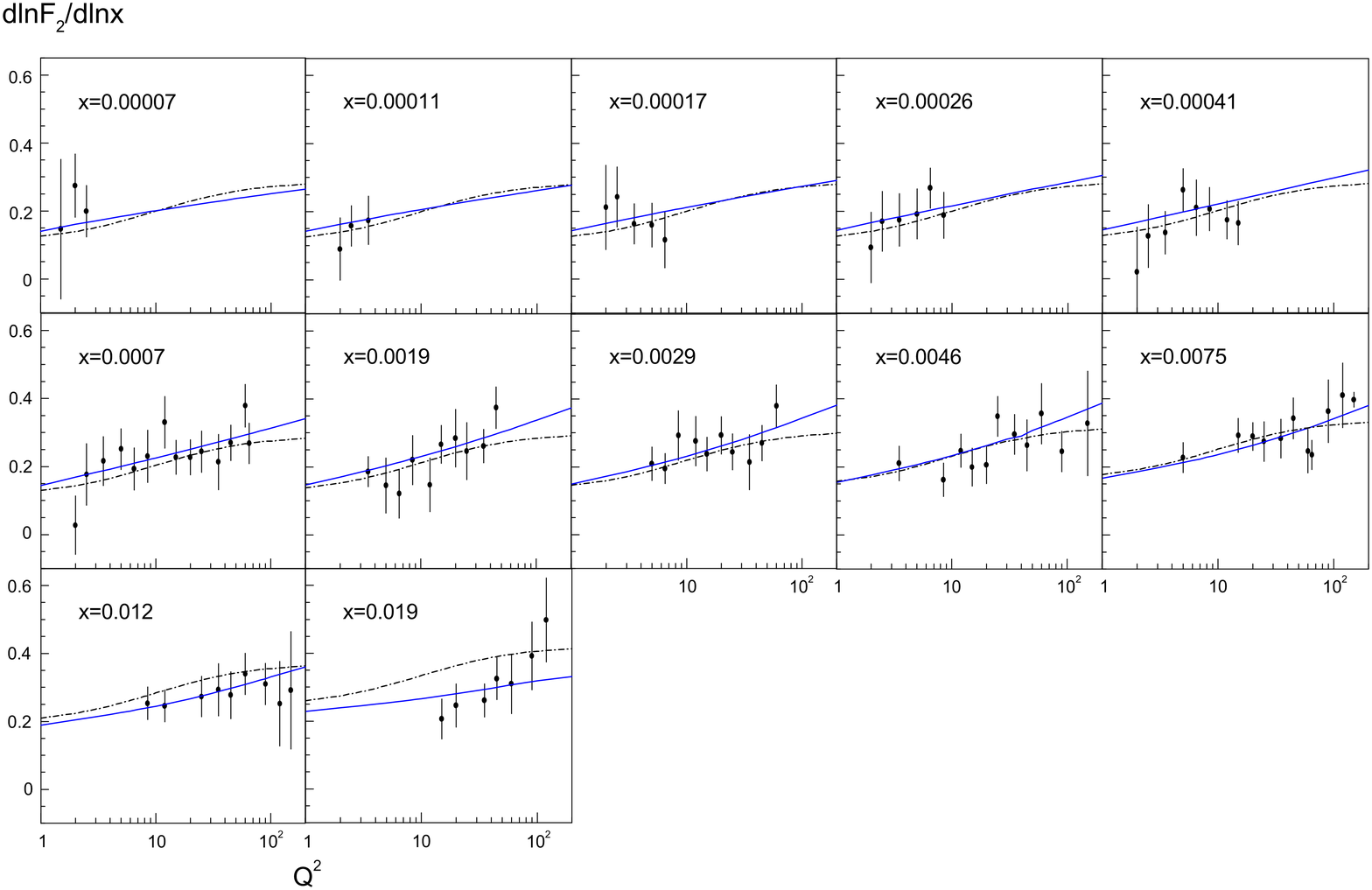} 
\caption{\small The result for the partial derivative $B_{x}$ slope 
plotted as a function of $Q^{2}$ for fixed $x$ compared with the 
latest H1 data \cite{H1c2}. Same notation of the previous plots.} 
\end{center} 
\end{figure} 
 
In Fig. (7) are shown the slope $B_{x}(x,Q^{2}_{fix})$ for virtualities 
ranging from $1.5 \leq Q^2 \leq 150$ GeV$^2$.  One verifies that the slope is 
independend of Bjorken $x$ for $x\leq 0.01$, in agreement with the
experimental  measurements and consistent with the H1 NLO QCD fit \cite{H1c2}.
The two  backgrounds deviate from  each one in the transition region $\approx
0.01$,  suggesting that the large $x$ region would be described differently
taking distinct  backgrounds. In a rough extrapolation, the CKMT-type seems to
be  favoured, whereas the  logarithmic-type would overestimate the large $x$
slope values.   
\begin{figure}[ht] 
\begin{center} 
\includegraphics*[scale=0.65]{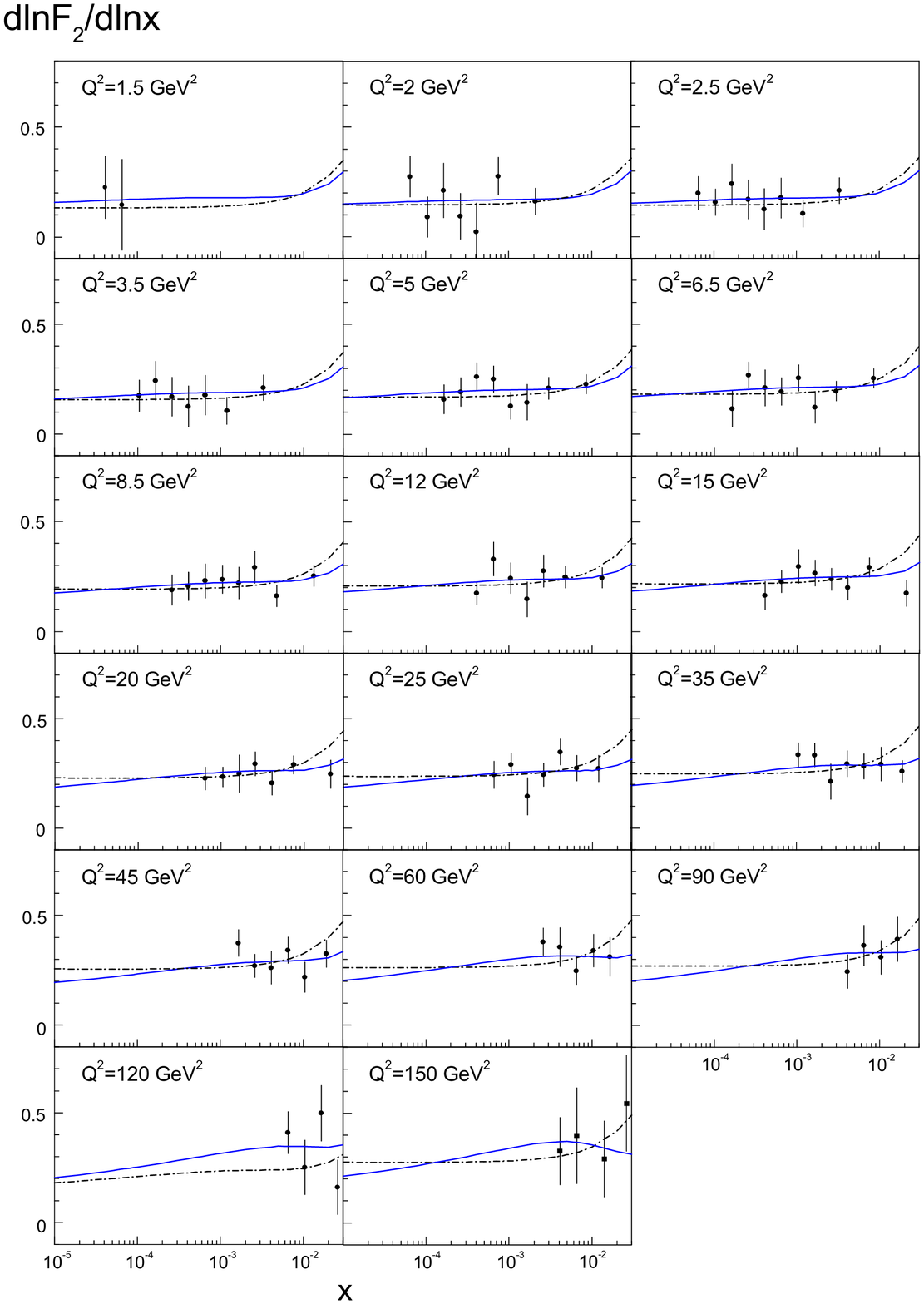} 
\caption{\small The result for the $B_{x}$ slope plotted as a function 
of $x$ for fixed $Q^{2}$ virtualities compared with the latest H1 
data \cite{H1c2}. Same notation of the previous plots.} 
\end{center} 
\end{figure}

\section{Conclusions}
When  looking in the  region of high energy limit, we are faced with
the lack of  connection between the Regge approach and perturbative QCD in
describing the asymptotic behavior of the hadronic (and photon initiated)
cross
 sections. The main issue is whether perturbation theory may shed light
on the
 origin and the nature of such physics, i.e the Pomeron induced
reactions. In
 this work we have studied in detail the application of the
finite sum of gluon
 ladders, associated with a  truncated BFKL series,  for
the inclusive structure
 function considering up to the two rung ladder
contribution. This 
 provides the asymptotic behavior expected from the
Froissart bound. The truncated
 series describes a broad region on $Q^2$,
however with a  formally large
 $\chi^2/\rm{dof}$ due to the
limited number of adjustable parameters. A large
 span in $Q^2$ is obtained if
we consider a non-perturbative background, modeled
 here as a soft Pomeron. The
resulting picture  is very close to the
 two-pomeron model of
Donnachie-Landshoff, with the hard Pomeron settled by the
 finite sum of
ladders. As a result, the structure function is described with
 good
confidence level for data lying at $x\leq 0.025$ and $0.045 \leq Q^2 \leq
1500$ GeV$^2$. The description of data with $x\geq 0.025$ is expected to
hold,
 since we have introduced the threshold factor, $(1-x)^7$.
There is
 no unambiguous sensitivity to the specific choice of the
background,
 although the CKMT-type is preferred due to the reduced number of
additional
 parameters. The region where the deviation between the backgrounds
is more
 important stays in the very low virtualities ($Q^2 \leq 1$ GeV$^2$),
with the
 logarithmic-type providing a better $\chi^2/\rm{dof}$.
 
Using the mentioned result, the  updated slopes
of the proton structure function can be numerically calculated considering the
two backgrounds. In general
 grounds, both choices for the non-perturbative
piece seem to be in agreement with the
 available data, with deviations in
kinematical regions where there are not
 measurements to clarify the analysis.
For example, the CKMT-type provides a description
 closer to the NLO QCD fits,
favoured due to the reduced number of
 additional constants. For instance, it
selects parameters leading to the growth with $Q^2$ at small $x$ for the
 slope
$(\partial F_2/\partial \ln Q^2)_{x}$,  and the flat behavior on small
Bjorken $x$ at fixed $Q^2$ for the slope  $(\partial \ln F_2/\partial \ln
x)_{Q^2}$, with similar features as in the QCD fits.

\section*{Acknowledgments}

MVTM thanks Prof. Peter Landshoff  for useful enlightenings on the two-pomeron
model.
 This work was partially supported by CNPq, Brazil.

\end{document}